\documentclass[aip,pop,reprint,superscriptaddress]{revtex4-1}
\usepackage{latexsym, amssymb, mathrsfs, graphicx}

\begin{document}
\title{Experimental evidence for collisional shock formation via two obliquely merging supersonic plasma jets}

\author{Elizabeth C.\ Merritt}
\email{emerritt@lanl.gov}
\affiliation{Los Alamos National Laboratory, Los Alamos, NM, 87545, USA}
\affiliation{University of New Mexico, Albuquerque, NM, 87131, USA}
\author{Auna L.\ Moser}
\author{Scott C.\ Hsu}
\email{scotthsu@lanl.gov}
\affiliation{Los Alamos National Laboratory, Los Alamos, NM, 87545, USA}
\author{Colin S.\ Adams}
\affiliation{Los Alamos National Laboratory, Los Alamos, NM, 87545, USA}
\affiliation{University of New Mexico, Albuquerque, NM, 87131, USA}
\author{John P. Dunn}
\author{A.\ Miguel Holgado}
\affiliation{Los Alamos National Laboratory, Los Alamos, NM, 87545, USA}
\author{Mark A.\ Gilmore}
\affiliation{University of New Mexico, Albuquerque, NM, 87131, USA}

\date{\today}

\begin{abstract}
We report spatially resolved measurements of the oblique merging of two supersonic laboratory plasma jets. 
The jets are formed and launched by pulsed-power-driven railguns using injected argon, and  have electron density $\sim 
10^{14}$~cm$^{-3}$, electron temperature $\approx 1.4$~eV, ionization fraction near unity, and velocity $\approx 
40$~km/s just prior to merging. The jet merging produces a few-cm-thick stagnation layer, as observed in both
fast-framing camera images and multi-chord interferometer data, consistent with collisional shock formation
[E. C. Merritt et al., Phys.\ Rev.\ Lett.~{\bf 111}, 085003 (2013)].
\end{abstract}


\maketitle 
\section{Introduction}

We have conducted experiments on the oblique merging of two supersonic plasma jets\cite{Merritt-prl13} on the Plasma 
Liner Experiment\cite{hsu-pop12} (PLX) at Los Alamos National Laboratory. These experiments were the 
second in a series of experiments intended to demonstrate the formation of imploding spherical plasma liners via an array 
of merging supersonic plasma jets.\cite{hsu-ieee12,cassibry12,cassibry13} The latter has been 
proposed\cite{thio99,thio01,hsu-ieee12} as a standoff compression driver for magneto-inertial
fusion\cite{lindemuth83,kirkpatrick95,lindemuth09} (MIF) and, in the case of targetless implosions, for
generating cm-, $\mu$s-, and Mbar-scale plasmas for high-energy-density physics\cite{drake} research. In our first 
set of experiments, the parameters and evolution of a single propagating plasma jet were characterized in 
detail.\cite{hsu-pop12} The next step beyond this work,
a thirty-jet experiment to form and assess spherically imploding plasma liners, has 
been designed\cite{hsu-ieee12,cassibry13,awe11} but not yet fielded.  A related jet-merging 
study\cite{case13,messer13,wu13} was also conducted recently by HyperV Technologies.

The supersonic jet-merging experiments reported here are also relevant to the basic study of plasma shocks\cite{jaffrin} 
in a semi- to fully collisional regime. Related studies include counter-streaming laser-produced plasmas supporting 
hohlraum design for indirect-drive inertial confinement fusion\cite{bosch,rancu,wan} and for studying astrophysically 
relevant shocks,\cite{woolsey,romagnani,kuramitsu,kugland,ross} colliding plasmas using wire-array Z
pinches,\cite{swadling13a,swadling13b} and applications such as pulsed laser deposition\cite{luna07} and laser-induced
breakdown spectroscopy.\cite{sanchez-ake} Primary issues of interest in these studies include the identification of shock 
formation, the formation of a stagnation layer\cite{hough09,hough10,yeates} between colliding plasmas, and the possible 
role of two-fluid and kinetic effects on plasma interpenetration.\cite{berger,pollaine,rambo94,rambo95}

In this paper we present detailed measurements of the stagnation layer that forms between two obliquely merging 
supersonic plasma jets in a semi- to fully collisional regime.  First, we briefly describe the experimental setup
(Sec.~\ref{sec:setup}).  Then we discuss observations of the stagnation layer emission morphology (Sec.~\ref{emission}) 
and density enhancements (Sec.~\ref{density85}). We also examine the observed stagnation layer thickness in the 
context of various estimated collision length scales and two-fluid plasma shock theory (Sec.~\ref{width}). Collectively, our 
observations are shown to be consistent with collisional shocks. We close with a discussion of the implications of 
our results on proposed imploding plasma liner formation experiments (Sec.~\ref{discussion}) and a summary
(Sec.~\ref{sec:summary}).

\section{Experimental setup}
\label{sec:setup}
\begin{figure}[!tb]
\begin{center}
\includegraphics[width=2.truein]{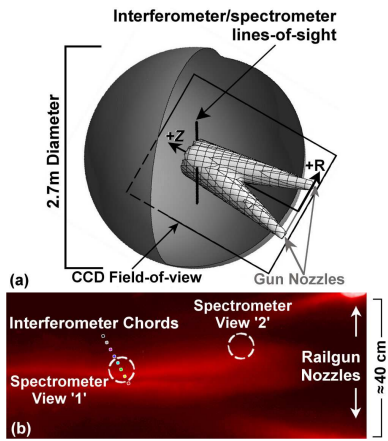}\\
\caption{(a) Schematic showing the spherical vacuum chamber, two merging plasma jets, ($R$,$Z$) coordinates used in 
the paper, approximate interferometer (representing all 8 chords) and spectrometer (view `1') lines-of-sight, and 
CCD 
camera field-of-view. (b) Location of interferometer chords (dots, $Z\approx 85$~cm, inter-chord spacing $= 1.5$~cm), 
and spectroscopy views (dashed circle, diameter $\approx 7$~cm) overlaid on a cropped CCD image of jet merging. 
Spectroscopy views `1' and `2' are located at $(R,Z) \approx (3.75~\mbox{cm}, 85~\mbox{cm})$ and  $(12~\mbox{cm}, 
55~\mbox{cm})$, respectively.}
\label{setup}
\end{center}
\end{figure}

Two plasma railguns are mounted on adjacent ports of a 2.7-m-diameter spherical vacuum chamber
[Fig.~\ref{setup}(a)], 
with a half-angle $\approx 12^\circ$ between the jet axes of propagation and a distance $\approx46$~cm between the
gun nozzles. At the nozzle exit, individual jets have initial parameters of peak electron density
$n_e\approx 2\times 
10^{16}$~cm$^{-3}$, peak electron temperature $T_e\approx 1.4$~eV, diameter $=5$~cm, and axial length $\approx 
20$~cm.\cite{hsu-pop12}  In this series of experiments, the initial jet velocity $V_{\rm jet}\approx 
40$~km/s and Mach number $M\equiv V_{\rm jet}/C_{\rm s,jet} > 10$, where $C_{\rm s,jet}$ is the sound speed in
the jet. 
More details on the railguns and the characterization of single-jet propagation are reported elsewhere.\cite{hsu-pop12} 
The jets are individually very highly collisional (thermal mean free paths $\lambda_{i} \sim  \lambda_{e}
\sim 100$~$\mu$m in a $\sim 20$-cm-scale plasma at initial jet merging), but the characteristic collision
length ($\sim 1$~cm, see Sec.~\ref{width}) 
between counter-propagating jet ions is on the order of the thickness of the observed stagnation layer that forms between the obliquely merging jets.

\begin{figure*}[!tb]
\begin{center}
\includegraphics[width=6.truein]{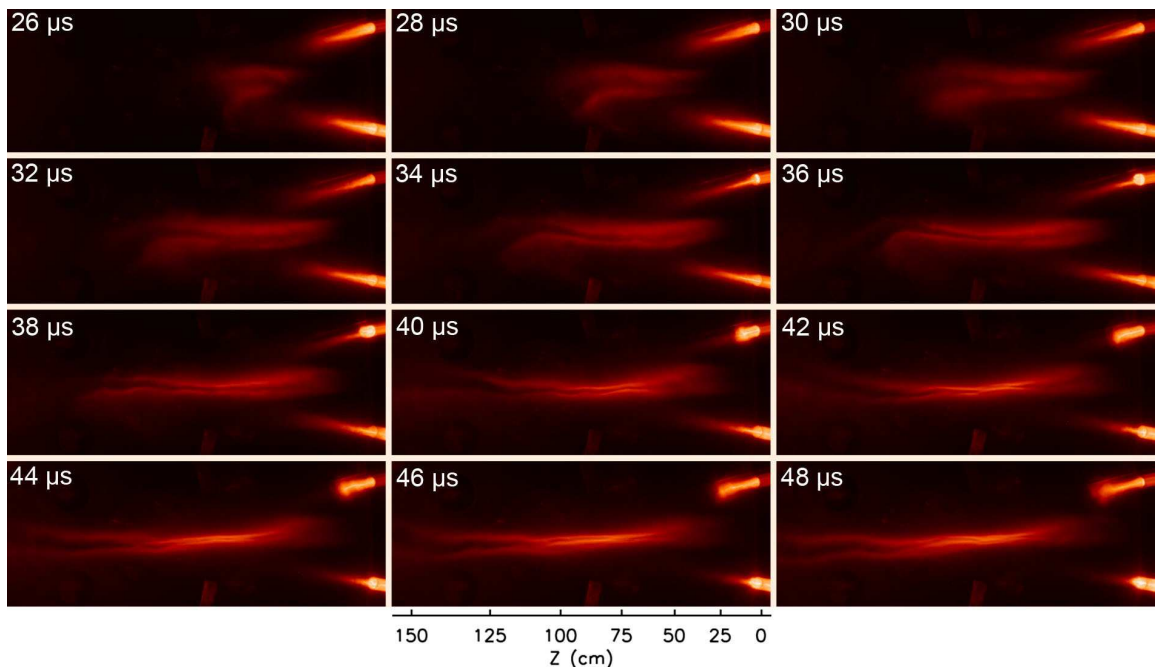}\\
\caption{False-color, cropped CCD images (log intensity, 20~ns exposure) of oblique jet merging [shots 1130, 1129, 1128, 
1127, 1125, 1122, 1120, 1132, 1134, 1136, 1138, 1140 (in order of timing)]. In each image, the two railgun nozzles 
($\approx 46$~cm apart) are visible on the right-hand-side, and the stagnation layer is oriented horizontally. All images 
have the same scale, which has a slightly nonlinear pixel-to-cm conversion due to the camera angle and optics.}
\label{pics}
\end{center}
\end{figure*}

The key diagnostics for our merging experiments are a visible-to-near-infrared survey spectrometer (0.275~m focal length 
with 600~lines/mm grating and 0.45~$\mu$s gating on the 1024-pixel microchannel plate array detector), an eight-chord 
561~nm laser interferometer,\cite{merritt-rsi12,merritt-htpd12} and an intensified charged-coupled-device (CCD)
visible-imaging camera (DiCam Pro, $1280\times1024$~pixels, 12-bit dynamic range). The CCD camera field-of-view 
extends 
from $Z \approx 0$--156~cm.  The interferometer chords
and spectrometer view `1' intersect the stagnation layer at $Z \approx 85$~cm [Fig.~\ref{setup}(b)], with an angle of $
\approx 36^\circ$ with respect to the jet-merging plane (into the page).
The line formed by the interferometer chords is roughly transverse to the stagnation layer ($\approx
30^\circ$ with respect to the $R$ direction), with inter-chord spacing of $\approx 1.5$~cm, spanning $R \approx 
0.75$--$11.25$~cm.
The $\approx 30^\circ$ angle with respect to $R$ introduces slight temporal offsets
($\approx 0.2$~$\mu$s between 
adjacent interferometer chords) for interferometer
data plots versus $R$.  The $\approx 36^\circ$ angle between the chords and the merge plane may lead to underestimates of plasma density enhancements and overestimates of local density minima due to the chords intersecting both shocked and unshocked plasma regions.
Spectrometer view `1' is centered on the interferometer chord at $(R,Z) = (3.75~\mbox{cm},
85~\mbox{cm})$. Spectrometer view `2' is located at $(R,Z) \approx (12~\mbox{cm}, 55~\mbox{cm})$ and
is oriented $\approx 31^\circ$ relative to the merge plane. The 
collimated spectrometer field-of-view has a divergence of $2.4^\circ$ and a diameter $\approx7 \pm0.5 $~cm 
at the measurement position.  Plasma jet velocity is determined via an array of intensified photodiode
detectors.\cite{hsu-pop12}
Figure~\ref{pics} shows a sequence of twelve CCD camera images (a different shot for each time; images are very reproducible) of the time evolution of jet merging and the formation of a stagnation layer along the
jet-merging
plane (midplane, horizontal in the images), with a double-peaked emission profile transverse ($R$ direction, 
vertical in the images) to the layer. Experiments were conducted with top jet only, bottom jet only, and both jets firing to 
enable the most direct  comparison between single- and merged-jet measurements. 

\begin{figure}[!tb]
\begin{center}
\includegraphics[width=2.1truein]{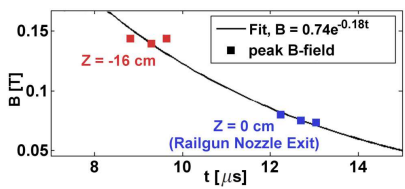}\\
\caption{Peak magnetic field (transverse to rails and jet propagation direction) vs.\ time (shots 2444, 2445, 2446, 2448, 
2449, 2450).  The data (squares) are from magnetic probes mounted at two positions along the exterior of the cylindrical railgun nozzle.}
\label{bfield}
\end{center}
\end{figure}

We have measured the jet magnetic field strength (transverse to the jet propagation direction) using magnetic probes mounted at two locations along the exterior of the cylindrical railgun nozzle.  The probe coils have nominal turns $\times$ area of 10~cm$^2$ (at the relevant frequency of 50~kHz, corresponding to the frequency of the gun current that produces the jet magnetic field), and the signals are passively integrated with a time constant of 0.322~ms. The jet is maintained at a constant diameter of 5~cm inside the nozzle. The field strength decreases from $\approx 0.14$~T at $Z=-16$~cm to $\approx 0.075$~T near the nozzle exit ($Z=0$~cm), with a decay time of 5.6~$\mu$s (see Fig.~\ref{bfield}). Extrapolating the decay to $t=24$~$\mu$s (i.e., $\approx 12$~$\mu$s after the jet  exits the nozzle), the field would be approximately 0.01~T. Based on the parameters at initial jet merging ($n_e = 2 \times 10^{14}$~cm$^{-3}$, $T_e = 1.4$~eV,\cite{hsu-pop12} $B = 0.01$~T and $v = 40$~km/s), then the ratio of the jet kinetic energy density ($\rho v^2/2$) to the magnetic energy density ($B^2/2\mu_0$) is 270.  The corresponding magnetic Reynold's number $R_m\approx 1.4$ (using a jet radial length scale of 5~cm for diffusion and a propagation distance 40~cm for advection), consistent with strong resistive field decay. If instead of being spatially uniform, the axial current producing the measured transverse magnetic field is peaked and mostly contained within a radius $r_0<r_{\rm nozzle}=2.5$~cm, then the peak field inside the jet would be larger than the measured value by a factor $B_0 r_{\rm nozzle}/r_0$, where $B_0=0.0035$~T.  If $r_0 = 1$~cm, then the peak $B=0.35$~T, which, extrapolating to $t=24$~$\mu$s, would give a kinetic-to-magnetic energy density ratio of $\approx 47$, still much larger than unity.  We also point out that the inferred decay time of 5.6~$\mu$s ignores jet expansion and cooling, meaning that 5.6~$\mu$s is an upper bound. Thus, we ignore magnetic field effects in this paper.  These magnetic field  measurements were taken during hydrogen experiments (the rest of the paper reports argon results), but $T_e\approx 1.4$~eV, and thus the magnetic diffusivity, were similar in both cases.

The argon plasma jets in these
experiments likely had high levels of impurities. The post-shot chamber pressure rise for 
gas-injection-only was about 30\% of that of a full railgun discharge, implying possible plasma impurity levels of 
up to 70\%. 
Identification of bright aluminum and oxygen spectral lines in our data\cite{Merritt-prl13} suggests that impurities are from 
the zirconium-toughened-alumina (0.15 ZrO$_2$ and 0.85 Al$_2$O$_3$) railgun insulators. Because the exact impurity 
fractions as a function of space and time in our jets are unknown, we bound our analysis by considering the two extreme 
cases of (i)~100\% argon and (ii)~30\% argon with 70\% impurities. For case (ii), we approximate the jet composition as 
43\% oxygen and 24\% aluminum (based on their ratio in zirconium-toughened-alumina) for spectroscopy analysis.

\section{Consistency of stagnation layer morphology with hydrodynamic shocks}\label{emission}

\subsection{Oblique shock morphology}

\begin{figure}[!tb]
\begin{center}
  \includegraphics[width=2truein]{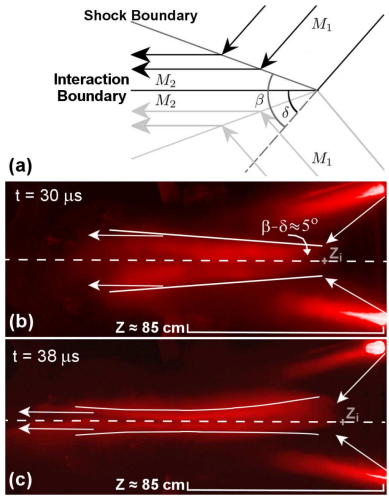}\\
  \caption{(a) Simple schematic of the interaction of two obliquely interacting supersonic flows with initial Mach numbers $M_1$. Flows are incident on the midplane with angle $\delta$. A shock boundary forms at angle $\beta$ with respect to the original flow direction. Post-shock flows have Mach number $M_2$ and flow direction parallel to the midplane. (b)~CCD image with postulated shock boundaries (solid white lines) and initial jet interaction distances $Z_i$ for shot 1128 at $t = 30$~$\mu$s ($Z_i\approx 30$~cm)
 and (c)~shot 1120 at $t = 38$~$\mu$s ($Z_i \approx 21$~cm). The field-of-view is the same for both CCD images.}
  \label{boundary}
  \end{center}
\end{figure}

 Because
inherently two-dimensional (2D) effects, such as non-uniform jet profiles, and time-dependent effects
do not permit a tractable analytic treatment of our problem and require full 2D simulations, we use analytic 1D
hydrodynamic theory to gain qualitative insight into the shock boundary morphology.  The assumption of
parallel, uniform flow within each jet [see Fig.~\ref{boundary}(a)]
reduces this to a 1D problem analogous to supersonic flow past a 
wedge or compression corner.\cite{landau,nunn} Comparing with the 1D theory, we show that the observed
emission layers (Fig.~\ref{pics}) are consistent with post-shocked plasma,\cite{Merritt-prl13}
with their edges (at larger $|R|$) corresponding to the shock boundaries.   

Figure~\ref{boundary}(a) shows a simple schematic of the jet interaction, where $\delta$ is the angle between the jet flow direction and the midplane, $M_1$ is the initial (pre-interaction) Mach number, and $\beta \equiv \beta(\delta, M_1)$ is the angle between the jet-flow direction and the position of an oblique shock boundary.\cite{nunn,drake} 
Figure~\ref{boundary}(b) shows a similar structure in a merged-jet CCD image. In this system, the turning angle $\delta \equiv \delta(Z_i)$ is given by $\tan \delta = (23 \mbox{ cm})/Z_i$, where $Z_i$ is the point at which the jets first interact, as determined by the appearance of emission [as indicated in Figs.~\ref{boundary}(b) and \ref{boundary}(c)]. The shock boundary angle $\beta$ is given by\cite{landau}
\begin{eqnarray}
\frac{23~\mbox{cm}}{Z_i} = 2 \cot \beta \left[ \frac{M_1^2 \sin^2 \beta - 1}{M_1^2( \gamma + \cos 2\beta) + 2} \right], 
\label{betaeqn}
\end{eqnarray}
and the opening angle of the shock relative to the midplane
is $\beta - \delta$. Determination of $Z_i$ from plasma emission
may slightly overestimate $Z_i$, but the errors introduced are small compared to the actual difference
between predicted and observed values of $\beta - \delta$ (presented below).
Also, a slight overestimate of $Z_i$ does not affect the discussion in Sec.~\ref{transition} regarding a possible
shock transition.

\begin{figure}[!tb]
\begin{center}
\includegraphics[width=2.truein]{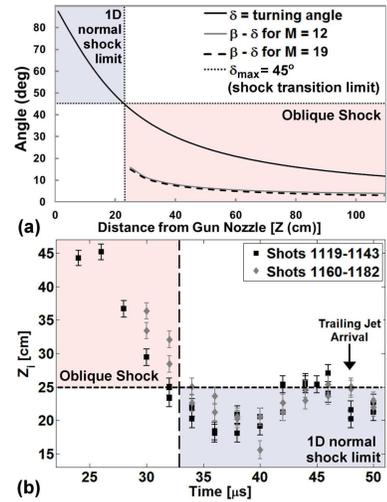}
\caption{(a) Plot of $\delta$ and $\beta - \delta$ vs.\ $Z$ for both $M = 12$ and $M=19$ (from 1D hydrodynamic
theory). The predicted threshold turning angle (also from the 1D theory),
$\delta = \delta_{max} = 45^\circ$, and corresponding $Z_i(\delta_{max}) \approx 25$~cm are marked with 
horizontal and vertical dotted lines, respectively. (b) Plot of $Z_i$ vs.\ time for shots 1119--1143 and shots 1160--1182. 
Error bars correspond to a $\pm 7.5$~pixel offset along the $Z$ axis during image processing. The 1D theoretical cutoff 
for oblique shock formation, $Z_i \approx 25$~cm, is indicated by a horizontal dashed line.}
\label{angles}
\end{center}
\end{figure}

Assuming $T_e = 1.4$~eV, mean charge $\bar{Z}  = 0.94$ (both inferred from spectroscopy at 
$Z = 41$~cm),\cite{hsu-pop12} and specific heat ratio $\gamma = 1.4$,\cite{awe11} then a 100\% argon plasma jet with 
$V_{\rm jet} = 40$~km/s has $M = 19$. For the 30\%/70\% mixture composition, $T_e = 1.4$~eV and $\bar{Z}  = 0.92$ 
(see Sec.~\ref{density85}), which are similar to the 100\% argon case. To place a stringent lower bound on $M$ for the 
30\%/70\% case, we use an ion-to-proton mass ratio $\mu = \mu_{O}=16$ because oxygen is the lightest element in the 
impurity mixture. Thus, we estimate that $12<M<19$.  We find that 
predicted $\beta-\delta$ values are very similar for $M=12$ and 
$M=19$ for a range of $Z$ [Fig.~\ref{angles}(a)].

We observe that $Z_i$ falls from $Z_i \approx 45$~cm at $t = 26$~$\mu$s to  $Z_i \approx 18$~cm at $t = 36$~$\mu$s 
[Fig.~\ref{angles}(b)], consistent with
jet axial expansion\cite{hsu-pop12} that reduces the velocity and thus increases the 
jet expansion angle for the rear portion of the jet.  A second dip in $Z_i$ beginning at $t \approx 47$~$\mu$s is due to 
the arrival of a trailing jet (created by ringing in the underdamped railgun current\cite{hsu-pop12}) at the merge region. 
For early times $t \approx 24$--$33$~$\mu$s ($Z_i \approx 45$--$25$~cm), the 1D theory predicts oblique shock 
formation consistent with the observed wedge-shaped emission boundary, as 
illustrated in Fig.~\ref{boundary}(b). In this case, the measured $\beta - \delta \approx 
5^\circ$. For $M = 12$--$19$, the theoretically predicted $\beta - \delta \approx 11^\circ$, which is within approximately a 
factor of two of the experimentally inferred value. This is reasonable agreement given that the 1D prediction does not 
include 2D/3D nor plasma equation-of-state\cite{messer13} (EOS) effects.

\subsection{Speculation on shock transition}
\label{transition}

There is a possible emission morphology transition between earlier and later times
[i.e., Fig.~\ref{boundary}(b) versus \ref{boundary}(c)].  For the theoretical 1D problem with uniform, parallel
flow [with an angle $\delta$ relative to the ``interaction boundary" in Fig.~\ref{boundary}(a)],
there is a threshold $\delta_{\rm max} \approx 45^\circ$ ($Z_i \approx 25$~cm)
beyond which no oblique shock forms.  In 2D theory, this corresponds to a detached shock, which we treat
in the 1D analysis by considering the limiting case of a normal shock.
Predicting the exact structure of a detached shock
in our 2D geometry, including spatial non-uniformities and
time-dependence, is not a tractable analytic problem and requires 2D simulations beyond the scope of this paper.
However, we can still compare our postulated morphology transition with $\delta_{\rm max}$ from the 1D theory.
As shown in Fig.~\ref{angles}(b), our observed $Z_i$ falls below (and hence $\delta$ rises above)
the transition threshold (predicted by the 1D theory)
around 33~$\mu$s, consistent with the approximate time of the possible
morphology transition between Figs.~\ref{boundary}(b) and \ref{boundary}(c).
While it is far from conclusive that our observations show a shock transition or a detached shock,
they are suggestive and motivate more detailed future work.

\section{Observation of merged-jet densities exceeding that of interpenetration}
\label{density85}

If the merged-jet emission layers are post-shocked plasma, then we expect an increase in density across the shock 
boundary during jet merging. The density increase across a 1D shock boundary should satisfy the 1D Rankine-Hugoniot 
relation\cite{drake}
\begin{eqnarray}
\frac{n_2}{n_1} = \frac{(M_1 \sin \beta)^2 (\gamma + 1)}{(M_1 \sin \beta)^2 (\gamma - 1) +2},
\end{eqnarray}
where $n_1$ and $n_2$ are the pre- and post-shock densities, respectively. We bound
the theoretically predicted density changes in the system by using $M_1 = 12$--$19$, as well as the $\beta$ range corresponding to the observed $Z_i(\delta$). 
We use the conservative value of $\gamma=1.4$, as suggested by
recent work in a similar parameter regime,\cite{awe11} as a simple way to model ionization
and EOS effects in the theoretical estimates of Mach number and density enhancement.
Because the $Z_i$ range encompasses $\delta > \delta_{\rm max}$, we consider the limiting case of detached shock formation (corresponding to a normal shock, i.e.,
$\beta=90^\circ$, in 1D theory) in addition to oblique shocks. For an oblique shock with $M = 12$, Eq.~(\ref{betaeqn}) gives $\beta = 34^\circ$--$59^\circ$ for measured $Z_i = 45$--$25$~cm. Thus, the range of $n_2/n_1 = 5.4$--$5.7$. Similarly, for $M = 19$ we find $\beta = 34^\circ $--$58^\circ$ and $n_2/n_1 = 5.7$--$5.9$. Assuming normal shocks ($\beta = 90^\circ$) for $Z_i <25$~cm, we find $n_2/n_1 = 5.8$--$5.9$ for $M = 12$--$19$. Thus, the overall range
across the shock boundary is $n_2/n_1 = 5.4$--5.9, according to hydrodynamic 1D theory. 

\begin{figure*}[!tb]
\begin{center}
  \includegraphics[width=4.5truein]{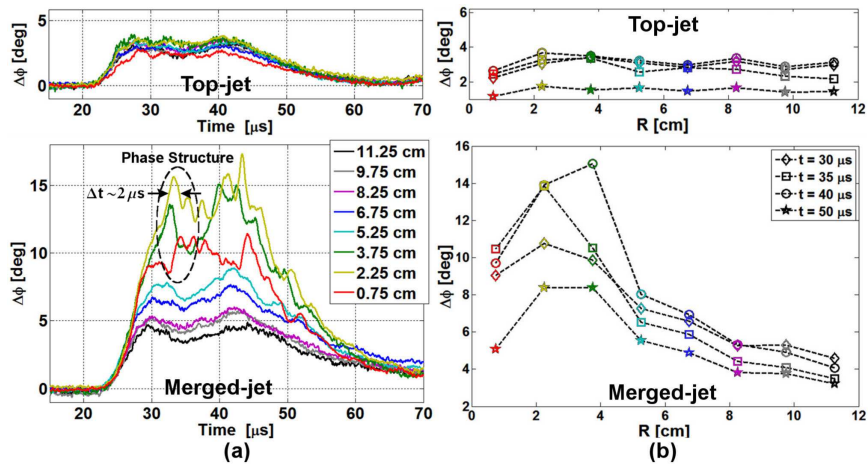}\\
  \caption{(a)~Phase shift vs.\ time at $Z\approx 85$~cm for top-jet-only (shot 1265) and merged-jet (shot 1120) cases. The merged-jet phase shift shows multiple small phase peaks of amplitude $\approx 2.5^\circ$  and width $\Delta t \approx 2$~$\mu$s. One such phase structure is highlighted by the dashed circle.  (b)~Phase shift vs.\ 
  interferometer chord position at several times for the same shots.}
  \label{phase}
  \end{center}
\end{figure*}

Next we compare $n_2/n_1$ from the 1D theory with the measured density enhancement 
$n_{\rm merged}/n_{\rm single}$ of the merged- over single-jet cases.
We calculate the ion-plus-neutral density $n_{\rm tot}$ using an interferometer
phase shift analysis accounting for multiple ionization states and the presence of impurities (see
Appendix~\ref{app:int}).
According to Eq.~(\ref{phase-err}), to determine $n_{\rm tot}$ we need the 
interferometer phase shift $\Delta \phi$, mean
charge $\bar{Z}$, the correction $Err$ [Eq.~(\ref{eq:err})] accounting for all non-free-electron contributions to $\Delta \phi$, and the 
interferometer chord path length (approximated by the plasma jet diameter).  The maximum correction $Err_{\rm max}$
is the largest scaled sensitivity, $C_{0,k}/C_e$ [Eq.~(\ref{eq:err-max})].  These are for Ar~\textsc{i}:~0.08 ($\delta N_n^{STP}= 2.8 \times 10^{-4}$ at $\lambda = 561$~nm, $\rho^{STP} = 1.6$~g/L),\cite{crc-opt,crc} for O~\textsc{i}:~0.03 ($K_{OI}m_{O} \approx 4.4 \times 10^{-24}$~cm$^3$ for $5000\mbox{ K} < T < 10000$~K),\cite{ivanova} and for Al~\textsc{i}:~0.007 ($\delta N_n^{STP} = 6.2 \times 10^{-2}$ at $\lambda = 561$~nm, $\rho^{STP} = 2.7$~g/cm$^3$).\cite{racik,crc} Thus, $Err_{\rm max}=0.08$ (for Ar~\textsc{i}).

\begin{figure}[!htb]
\begin{center}
\includegraphics[width=2.3truein]{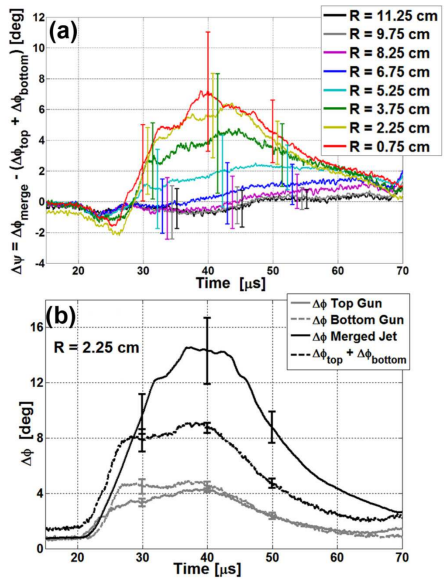}\\
\caption{(a) The difference between merged-jet and the sum of single-jet phase shifts $\Delta\psi$ vs.\ time for data 
averaged over shots 1117--1196 (merged-jet), shots 1277--1278 (bottom-jet) and shots 1265--1267 (top-jet). (b) Multi-shot 
(same data sets) averaged interferometer phase shift vs.\ time at  $R = 2.25$~cm, for top-, bottom-, and merged-jet 
cases. Error bars indicate the standard deviation of $\Delta\psi$ or $\Delta\phi$ over the stated data set.}
\label{compare}
\end{center}
\end{figure}

First, we determine $\Delta \phi_{\rm single}$ and $\Delta\phi_{\rm merged}$ for the single- and merged-jet cases, respectively.  The single-jet peak $\Delta\phi_{\rm single}$, averaged across chords for a single shot, 
is $\Delta\phi_{\rm single} \approx 4.0^\circ \pm 0.6^\circ$, where $0.6^\circ$ is the standard deviation [Fig.~\ref{phase}(a)]. The peak $\Delta\phi_{\rm single}$ averaged over multiple top-jet-only shots at the $R=2.25$~cm chord is $\Delta \phi_{\rm single} =4.3^\circ \pm 0.3^\circ$ [Fig.~\ref{compare}(b)], and thus we assume $\Delta\phi_{\rm single}=4^\circ$ for evaluating $n_{\rm tot,single}\equiv n_{\rm single}$. Merged-jet $\Delta\phi_{\rm merged}$ traces for a single shot show [Fig.~\ref{phase}(b)] a non-uniform spatial profile with a peak near the midplane and peak magnitude $\Delta\phi \approx 14^\circ$. At $R = 2.25$~cm, the peak $\Delta \phi = 14.3 \pm 2.4^\circ$ averaged over multiple shots [Fig.~\ref{compare}(b)]. Thus, we assume $\Delta\phi_{\rm merged}=14^\circ$ for evaluating $n_{\rm tot,merged}\equiv n_{\rm merged}$. 

Before evaluating $n_{\rm single}$ and $n_{\rm merged}$, we examine
$\Delta\phi$ enhancements for merged- over single-jet experiments by considering the quantity
\begin{equation}
\Delta\psi = \Delta\phi_{\rm merged} - (\Delta\phi_{\rm top} + \Delta\phi_{\rm bottom}),
\end{equation}
where $\Delta\phi_{\rm top}$ and $\Delta\phi_{\rm bottom}$ are
from top-jet-only and bottom-jet-only shots, respectively.
We use $\Delta \phi$ values averaged over multiple shots (Fig.~\ref{compare})
to reduce potential errors introduced by shot-to-shot variations.
A $\Delta \psi > 0$ implies a density of the merged-jet beyond that of the sum of single
jets and/or an increase in $\bar{Z}$ over that of a single jet.
Merged-jet measurements over the data set considered (merged-jet: shots 1117--1196; bottom-jet: shots 1277--1278;
top-jet: 
shots 1265--1267) show that $\Delta \psi > 0$ for $R \le 5.25$~cm [Fig.~\ref{compare}(a)], implying that simple jet 
interpenetration cannot account for the observed stagnation layer $\Delta\phi_{\rm merged}$. For $R \ge 6.75$~cm, 
$\Delta \psi$ is small because this region is outside the stagnation layer.

Now we evaluate $n_{\rm single}$ and $n_{\rm merged}$ in order to estimate the density enhancement
$n_{\rm single}/n_{\rm merged}$ at $Z\approx 85$~cm and $R\le 5.25$~cm. 
Having determined $\Delta\phi$ and $Err_{\rm max}$, we need only $\bar{Z}$ to estimate $n_{\rm single}$
and $n_{\rm merged}$.  The $T_e$ and $\bar{Z}$ are determined by comparing
spectral data\cite{hsu-pop12,Merritt-prl13}
with non-local-thermodynamic-equilibrium (non-LTE) spectral calculations 
in the optically thin limit using PrismSPECT\@.\cite{macfarlane03}
To mitigate the impact of line-of-sight effects
on our spectral analysis, we used the appearance (e.g., Ar~\textsc{ii}) and absence (e.g., impurity
Al~\textsc{iii}) of spectral
lines in the data (typically varying only in intensity) in the time range of interest to determine bounds on 
peak $T_e$ and $\bar{Z}$.
A single jet (assuming 100\% Ar) has a jet 
diameter $\approx 22$~cm at $Z \approx 80$~cm and $\bar{Z} $ = 0.94 ($T_e=1.4$~eV)
at $Z\approx 41$~cm (the emission is too low at $Z = 85$~cm to
infer $\bar{Z}$ there).\cite{hsu-pop12} Using $\bar{Z}=0.94$ with $\Delta\phi_{\rm single}=4.0^\circ$, we obtain $n_{\rm single}=2.1$--$2.3 \times 
10^{14}$~cm$^{-3}$ (bounds provided by $Err=0$ and $Err_{\rm max}=0.08$). 
For the 30\%/70\% mixture case (at the same $T_e=1.4$~eV),
$\bar{Z}=0.92$,\cite{Merritt-prl13} and therefore the $n_{\rm single}$ estimate changes by only a few percent.

To infer $\bar{Z} $ and $T_e$ for the merged-jet case, and therefore $n_{\rm merged}$, at $Z\approx 85$~cm,
we examine spectral data from spectrometer view `1.'
For 100\% argon, we infer 
that peak $T_e \ge 1.4$~eV and $\bar{Z}  \ge 0.94$.\cite{Merritt-prl13}
For the 30\%/70\% mixture, we infer that $2.2$~eV$\le$ peak $T_e < 2.3$~eV and
$1.3\le \bar{Z} < 1.4$, with the upper bounds determined by the absence of an Al~\textsc{iii} line in the
data.\cite{Merritt-prl13} 
Thus, for the 100\% argon case, we see little-to-no change in $\bar{Z}$ compared to the 
single-jet measurements, but the 30\%/70\% mixture calculation predicts an increase in $\bar{Z}$ during jet 
merging, accounting for some of the observed $\Delta\phi$ enhancement.
Using $\Delta \phi = 14^\circ$, chord path length of 22~cm, and $\bar{Z} $ = 0.94 (100\% argon case), 
we obtain $n_{\rm merged}= 
7.5$--$8.2 \times 10^{14}$~cm$^{-3}$
(bounds provided by $Err=0$ and $Err_{\rm max}=0.08$). In this case the density increase $n_{\rm merged}/n_{\rm single} = 3.2$--3.8. For the most conservative $\bar{Z}  = 
1.4$ of the 30\%/70\% mixture case, $n_{\rm merged}= 5.0$--$5.3 \times
10^{14}$~cm$^{-3}$, and $n_{\rm merged}/n_{\rm single} = 2.1$--2.4.  These values are summarized
in Table~\ref{ntable}. 

The observed range 
of $n_{\rm merged}/n_{\rm single} = 2.1$--3.8 exceeds the factor of two expected for jet interpenetration, although it is 
smaller than the $n_2/n_1=n_{\rm shock}/n_{\rm unshocked} = 5.4$--5.9 predicted by 1D theory.
Note that plasma diameter enhancement (along the interferometer chord direction)
in the merged- over the single-jet case, which we have not characterized,
and overestimates of $\bar{Z}$ (given that
we do not have a direct measurement at $Z\approx 85$~cm) would both lead to reductions
in our estimate of $n_{\rm merged}/n_{\rm single}$.  
The difference between the measured and predicted 
density jumps could be due to 3D (e.g., pressure-relief in the out-of-page dimension) and/or plasma
EOS effects not modeled by 1D hydrodynamic theory.  

\begin{table}[!b]
  \centering
\begin{tabular}{c c c}
  \hline \hline
   & 100\% Ar & 30\%/70\% \\ \hline
  $T_{\rm e,merged}$ & $\ge 1.4$~eV  & 2.2~eV$\le T_e <$2.3~eV \\
  $\bar{Z}_{\rm single}$ & 0.94 & 0.92 \\
  $\bar{Z}_{\rm merged}$ & 0.94 & 1.4 \\
  $n_{\rm single}$ & 2.1--2.3$\times 10^{14}$~cm$^{-3}$ & 2.2--2.4$\times 10^{14}$~cm$^{-3}$ \\
  $n_{\rm merged}$ & 7.5--8.2$\times 10^{14}$~cm$^{-3}$ & 5.0--5.3$\times 10^{14}$~cm$^{-3}$ \\
  $n_{\rm single}/n_{\rm merged}$ & 3.2--3.8 & 2.1--2.4 \\
  \hline \hline
\end{tabular}
\caption{Summary of the experimentally inferred
jet density enhancement at $Z \approx 85$~cm 
for the two mixture cases: 100\%~Ar and 30\%~Ar/70\%~impurities. Single-jet and merged-jet densities are calculated 
using $\Delta\phi = 4^\circ$ and $\Delta\phi = 14^\circ$, respectively, jet diameter of 22~cm, and $Err_{\rm max} = 0.08$.
Note that values for $\bar{Z}_{\rm single}$ are from $Z\approx 41$~cm.\cite{hsu-pop12}}
\label{ntable}
\end{table}

We point out a few additional features from the interferometry.
The spatial profile for the merged-jet $\Delta\phi$, as seen in Fig.~\ref{phase}(b), is peaked a few centimeters
away from the midplane ($R=0$) and correlates with the peaked emission profile in the $R$ direction,
as seen in the CCD images (Fig.~\ref{pics}).
Figure~\ref{phase}(a) shows evidence of variations in $\Delta \phi_{\rm peak} \approx 2.5^\circ$
over $\Delta t \approx 2$~$\mu$s in the merged-jet measurements that are not present in single-jet experiments. 
Assuming $V_{\rm jet}=40$~km/s, the width of the indicated structure is $\approx 8$~cm. The appearance of this $\Delta
\phi$ structure alternates between adjacent chords for chords at $R = 0.75$--3.75~cm, i.e., the 
$\Delta\phi$ rise in one chord corresponds to a fall in another chord at $\approx 1.5$~$\mu$s intervals. Because the 
inter-chord distance is 1.5~cm, the structure has a transverse velocity $\approx 15$~km/s. The underlying cause of these 
structures has not yet been determined.

\begin{figure}[!tb]
\begin{center}
\includegraphics[width=3.3truein]{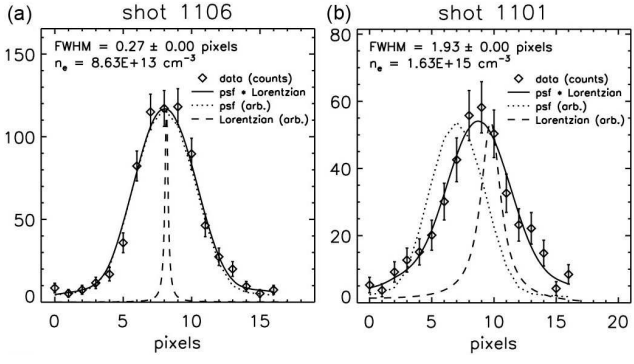}\\
\caption{Determination of electron density $n_e$ at $t = 30$~$\mu$s via Stark broadening of the H-$\beta$ lines for (a) 
top-jet-only (shot 1106) and (b) merged-jet (shot 1101) cases at the spectrometer position `2' [see Fig.~\ref{setup}(b)].
Shown are the experimental data (diamonds with error bars $\pm \sqrt{\rm counts}$), an overlay of the measured 
instrumental broadening profile (dotted line, labeled as `psf' for point spread function), and a Lorentzian H-$\beta$ profile 
(dashed line) that gives the best fit (minimum $\chi^2$) of the convolution (solid line) of the psf and the Lorentzian to the 
data.}
\label{stark}
\end{center}
\end{figure}

Electron density results (determined via Stark broadening of the H-$\beta$ line) at spectrometer view `2' 
($Z\approx 55$~cm)
also show a density enhancement: from $n_e \le 8.6 \times 10^{13}$~cm$^{-3}$ (shot 1106) for a top-jet-only case to 
$n_e \approx 1.6 \times 10^{15}$~cm$^{-3}$ (shot 1101) during jet merging (Fig.~\ref{stark}).  The electron density was 
determined via\cite{hsu-pop12}
\begin{eqnarray}
n_e = 6.05 \times 10^{14} [\mbox{FWHM}(\mbox{pixels})]^{3/2} \mbox{ cm}^{-3},
\end{eqnarray}
where FWHM is the full-width-half-maximum of the Stark-broadened H-$\beta$ line (more details given in the caption for 
Fig.~\ref{stark}). For the top-jet-only shot (1106), the FWHM of the Lorentzian (with instrumental broadening removed) is 0.27~pixels, which is significantly less than the 1 pixel spectrometer resolution. So, we consider $n_e = 8.6 \times 10^{13}$~cm$^{-3}$ an upper bound, i.e., the density could be less but is too small to be resolvable. Thus, $n_{\rm e,merged}/n_{\rm e,single} \gtrsim 10$ at $Z \approx 55$~cm, which is significantly larger than the $n_{\rm merged}/n_{\rm single}$ observed at $Z \approx 85$~cm. Some of the $n_e=\bar{Z}n_{\rm tot}$ increase is likely due to increased ionization during jet merging, but unfortunately there was not enough information in the measured spectrum at $Z\approx 55$~cm to infer $\bar{Z}$. The $Z\approx 55$~cm measurements were taken at a larger distance from the jet axes than the $Z\approx 85$~cm measurements, which, along with possibly a different $\bar{Z}$ at $Z\approx 55$~cm, could contribute to the difference in density enhancements observed at the two different locations. Nevertheless, the magnitude of the $n_e$ enhancement suggests the presence of post-shocked density also at $Z\approx 55$~cm.

\section{Collisionality estimates and comparison to two-fluid plasma shock theory}
\label{width}

Both the experimentally measured
emission\cite{Merritt-prl13} and interferometer $\Delta \phi$ [Fig.~\ref{phase}(b)]  have the same gradient length 
scale (few cm) in the $R$ direction,
and the $\Delta \phi$ dip at $R = 0.75$~cm and peak at $R = 2.25$--$3.75$~cm [Fig.~\ref{phase}(b)]
are well-aligned with the emission dip and peak, respectively.\cite{Merritt-prl13}
In this section, we compare these observations
with the expected scale sizes of collisional plasma shock formation via colliding plasmas.  
For the latter, the stagnation layer thickness is expected\cite{rambo94} to be on the order of the
ion penetration length into the opposing jet.
We find that, in our parameter regime, the limiting physics for ion penetration is frictional drag exerted by the
ions of one jet on the counter-streaming ions of the other jet.  This is evaluated using the slowing-down rate in the fast approximation,\cite{nrl-formulary}
\begin{equation}
\nu^s_{ii^\prime} = 9.0\times10^{-8} n_i^\prime Z^2 Z^{\prime 2} \ln \Lambda \left(\frac{1}{\mu} + \frac{1}{\mu^\prime}
\right)\frac{\mu^{1/2}}{\epsilon^{3/2}},
\end{equation}
where (see Appendix~\ref{app:logL})
\begin{equation}
\ln \Lambda=43- \ln \left[ \frac{Z Z^\prime (\mu + \mu^\prime)}{\mu \mu^\prime (v_{\rm rel}/c)^2}
\left(\frac{n_e}{T_e}\right)^{1/2}\right]
\label{logL-text}
\end{equation}
is the Coulomb logarithm for counter-streaming ions (with relative velocity $v_{\rm rel}$) in the presence of
warm electrons,\cite{nrl-formulary} $n_i$ and $n_e$
[cm$^{-3}$] the ion and electron densities, respectively, $Z$ the mean charge state, $T_e$ [eV] the electron temperature,
$\epsilon$ [eV] the relative kinetic 
energy of the test particle, $c$ the speed of light,
and the unprimed and primed variables correspond to a test particle from one jet and the field particles of the other jet, respectively.  The ion penetration length is
\begin{equation}
\lambda^s_i \approx \frac{v_{\rm rel}}{4 \sum_{i^\prime} \nu_{i i^\prime}^s},
\end{equation}
where the factor of 4
results from the integral effect of $v_{\rm rel}$ slowing down to zero,\cite{messer13} and the
summation is over all field-ion species for the mixed-species jet case.
We estimate $\lambda^s_i$ by considering jets of 100\% argon and the 30\%/70\% mixture (specifically, 30\%~Ar, 43\%~O, 24\%~Al), in all cases using $v_{\rm rel} = 20$~km/s 
(corresponding to $\delta = 30^\circ$ and $Z_i \approx 40$~cm)
and the plasma parameters listed in
Table~II, which also contains a summary of the ion-electron slowing-down distances $\lambda_{ie}^s$
calculated
using the slow approximation for $\nu_{ie}^s$ and the Coulomb logarithm for
ion-electron collisions.\cite{nrl-formulary}
For inter-species collisions between mixed-species jets (due to impurities),
we use $n_i = (\mbox{\% ion species}) \times n_{\rm tot}$.  

\begin{table}[!tb]
  \centering
\begin{tabular}{lcccc}
  \hline \hline
   &  & 100\% Ar & & 30\%/70\% mixture \\ \hline
   $n_{\rm tot}$ (cm$^{-3}$) & & $8 \times 10^{14}$ & & $5 \times 10^{14}$ \\
   $T_e$ (eV) & & 1.4 & & 2.2\\
   $\bar{Z} $ & Ar & 0.94 & & 1.2 \\
   & Al & & & 2.0\\
   & O & & & 1.0\\ \hline
  $\lambda^s_i$ (cm) & Ar & 1.8 & & 0.8 \\
   & Al & & & 0.2\\
   & O & & & 0.3\\
  $\lambda^s_{ie}$ (cm) & Ar & 17.3 & & 25.1 \\
   & Al &  & & 6.6\\
   & O &  & & 14.1\\
  \hline \hline
\end{tabular}
\label{lambda-table}
\caption{Summary of stopping lengths for inter-jet particle interactions, for both the 100\% Ar and 30\%/70\% mixture 
cases.}
\end{table}

We also estimate the inter-jet mean free path (mfp) of Ar$^{1+}$-Ar charge and momentum transfer. 
The assumption of $v_{\rm rel} = 20$~km/s gives a kinetic energy of $\approx 80$~eV, corresponding to charge and 
momentum transfer cross-sections $\sigma_{CT} \approx 0.3 \times 10^{-18}$~m$^{2}$ and $\sigma_{m} \approx 0.7 
\times 10^{-18}$~m$^{2}$, respectively.\cite{phelps90} The total mfp for Ar$^{1+}$-Ar interaction is $\lambda_{in} = 
1/\sigma_{\rm tot}  n_n = 1/[(\sigma_{CT} + \sigma_m) n_n]$, where $n_n = (1-\bar{Z} ) n_{\rm tot}$ (for $\bar{Z} <1$) is 
the neutral density. For the pure-argon merged-jet parameters (an upper bound on $n_n$ because $n_n/n_{\rm tot}
< 10^{-2}$ for the mixture case), $\lambda_{in} \approx 2$~cm $\gtrsim  \lambda^s_i$.
Comparing all these length scale estimates with the observed few-cm-thick stagnation layer implies
that our inter-jet merging is in a semi- to fully collisional regime.

Previously, we showed that the transverse ($R$) dynamics of our oblique jet merging compared favorably with 1D 
collisional multi-fluid plasma simulations of our experiment.\cite{Merritt-prl13} Specifically, reflected shocks
in the simulation (propagating in the $R$ direction) gave rise to a
double-peaked density profile (at $\pm R$) consistent with our density and emission profile measurements.
Here, we consider our experimental observations in the context of two-fluid plasma shock theory.\cite{jaffrin} In the 
case of a high-$M$, two-fluid shock, differing ion and electron transport results in shock structures on multiple spatial 
scales.\cite{jaffrin} The length scale of ion viscosity and thermal conduction effects is on the order of the collisional mfp 
of the shocked ions, $\lambda_i = v_{th,i}/\nu_i$, where $v_{th,i}$ and $\nu_i$ are the ion thermal velocity and thermal 
collision frequency, respectively, while the length scale of electron viscosity and thermal conduction effects is on the 
order of $\lambda_i \sqrt{m_i/m_e}$.\cite{jaffrin} The downstream mfp in our system is estimated to be on the order of 
$8\times 10^{-3}$~cm based on the merged-jet parameters given in Table~\ref{ntable}.   In order to bound the range of 
electron shock scale lengths, we use the limiting cases of $\mu = \mu_O=16$  and $\mu =\mu_{Ar}=40$, and obtain 
$\lambda_i \sqrt{m_i/m_e}\approx 1.4$--2.2~cm, which is of the same order as the gradient scale lengths of the 
observed emission\cite{Merritt-prl13} and $\Delta\phi$ profiles [Fig.~\ref{phase}(b)]. This suggests that our 
observations are 
also consistent with collisional two-fluid plasma shocks in that the observed scales could be large enough to contain
an electron-scale pre-shock.

\section{On the use of merging plasma jets for forming spherically imploding plasma liners}
\label{discussion}

A key motivation for this work was to study two obliquely merging supersonic plasma jets as the ``unit physics" process underlying the use of an array of such jets to form spherically imploding plasma liners.  The latter is envisioned as a standoff driver for MIF.\cite{thio99,thio01,cassibry09,hsu-ieee12,hsu-pop12,santarius12} The dynamics arising in the jet merging, e.g., shock formation, sets the properties of the subsequent, merged plasma that ultimately determines the liner uniformity and peak ram pressure ($\rho v^2$).  These physics issues have been considered recently via theory and numerical modeling.\cite{parks08,cassibry12,kim13} In spherical plasma liner formation via an array of plasma jets, the initial merging would be among more than two jets, and the detailed merging geometry would depend on the port geometry of the vacuum chamber. In the case of PLX, a quasi-spherical arrangement of 60 plasma guns would result in twelve groups of five jets, with each group arranged in a pentagonal pattern.

\begin{figure}
\includegraphics[width=2truein]{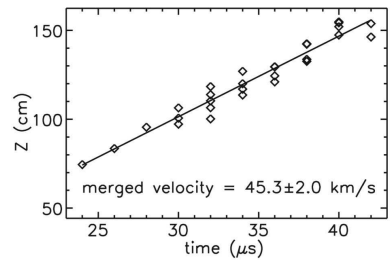}%
\caption{\label{fig:merged-speed} Leading edge position of the merged jet, as determined visually from CCD images, 
versus time (shots 1120--1172).  Diamonds are data points, and the black line is a linear fit giving the velocity
of the merged-jet leading edge.}
\end{figure}

A key figure of merit for implosion performance is the jet/liner Mach number $M$, i.e., a lower $M$ results in faster plasma spreading, density reduction, and lower ram pressure.\cite{parks08,awe11,davis12,cassibry13} A concern is that jet merging would lead to shock formation and heating that would significantly decrease $M$ (compared to its initial value) and, thus, implosion performance. The results reported here are encouraging in that the experimentally inferred increases in $T_e$ [by up to a factor of $(2.3~{\rm eV})/(1.4~{\rm eV})=1.64$] and $\bar{Z}$ (by up to a factor of $1.4/0.94=1.49$) lead to an increase in $C_s\sim (\bar{Z}T_e)^{1/2}$ of 56\% (we caution that
more data is needed to establish a more accurate upper bound on $T_e$ in the merged case).    We estimate the speed of the leading edge of the merged plasma to be $\approx 45$~km/s (see Fig.~\ref{fig:merged-speed}), which is close
to the initial jet speed of $\approx 41$~km/s.  An unchanged velocity after jet merging would result in a
modest 36\% reduction in $M$\@.

With regard to uniformity, the outstanding questions are how the observed structure in two-jet merging
would affect the uniformity of the leading edge of an imploding spherical plasma liner formed by multiple merging jets,
and how much non-uniformity would be tolerable for compression of a magnetized plasma target for application
to MIF\@.
This problem has been studied recently in two simulation studies,\cite{cassibry12,kim13} which reached opposing 
conclusions using two different codes employing very different numerical models and techniques.  One study
concluded that a series of shocks occurring during plasma liner convergence would degrade the implosion 
performance,\cite{kim13} while the other showed that initial non-uniformities arising from jet merging were
largely smeared out by the time of peak compression.\cite{cassibry12}  More detailed studies are needed to resolve the
discrepancy.  We envision a five-jet experiment on PLX followed by a 30- or 60-jet experiment to study this 
and other issues.

\section{Summary}
\label{sec:summary}
We have made spatially resolved measurements, in a semi- to fully collisional regime, of the stagnation layer 
that forms between two obliquely merging supersonic plasma jets. CCD images show a double-peaked emission profile 
transverse to the layer, with the central emission dip consistent with a density dip observed in the interferometer data. The 
stagnation layer thickness is a few cm, which is of the same order as the ion penetration length (in our case determined 
by frictional drag between counter-streaming ions). The observed stagnation layer emission morphology
shortly after jet merging is consistent with hydrodynamic oblique shock theory. The density increase from that of an 
individual jet to the density of the post-merge stagnation layer is greater than that of interpenetration, even accounting for 
the higher ionization estimates found for the high-impurity versus pure-argon analysis limits. The measured density 
increase is low compared to 1D theoretical hydrodynamic predictions, but discrepancies are expected due to multi-
dimensional and plasma EOS effects in the experiment. We did not observe a strong rise in $T_e$ or $\bar{Z}$, which, 
coupled with little observed change in the jet velocity after merging, is encouraging for proposed plasma liner formation 
experiments.

\begin{acknowledgments}
Significant portions of this work are from E. C. Merritt's doctoral dissertation.  We acknowledge HyperV Technologies 
Corp.\ for extensive advice on railgun operation, T.~P.\ Intrator and G. A. Wurden for sharing  
laboratory and diagnostic hardware, and J. T. Cassibry,
J. Loverich, and C. Thoma for useful discussions. This work was supported by the U.S. Dept.\ of Energy.
\end{acknowledgments}

\appendix

\section{Interferometer phase shift analysis}
\label{app:int}
Previous interferometer phase shift analysis \cite{merritt-htpd12} for this experiment assumed a singly ionized argon 
plasma, which was adequate for our single-jet experiments.\cite{hsu-pop12} In these two-jet merging experiments, the 
observation of higher ionization states and significant impurity percentages required generalization of the phase shift 
analysis.

For a plasma with multiple gas species and ionization states, we can write $\Delta\phi$ as a superposition of the  
contributions from the electrons and all possible ionization states for each gas species in the plasma:
\begin{eqnarray}
\Delta\phi_{\rm tot} &=& \Delta\phi_e - \sum_{j,k} \Delta\phi_{j,k} \\
&=& \int C_e n_e dl - \int \sum_{j,k} C_{j,k} n_{j,k} dl,
\end{eqnarray}
where $C_e$ is the interferometer sensitivity constant for electrons
and $C_{j,k}$ is the sensitivity constant for the $j$th ionization state ($j=0$ denotes neutrals)
of the $k$th gas species.

For a species with ionization state $j$, the electron density due to that species is $n_{e,j} = j n_j$. The total electron 
density is then $n_e = \sum_{j,k} n_{e,(j,k)} = \sum_{j,k} j n_{j,k}$. The average ionization state of the plasma is then
\begin{eqnarray}
\bar{Z}  = \frac{n_e}{n_{\rm tot}} = \frac{\sum_{j,k} n_{e,(j,k)}}{\sum_{j,k} n_{j,k}} = \frac{\sum_{j,k} j n_{j,k}}{\sum_{j,k}
n_{j,k}},
\end{eqnarray}
where $n_{\rm tot} = \sum_{j,k} n_{j,k}$ is the total ion-plus-neutral
density of the plasma. The phase shift equation becomes
\begin{eqnarray}
\Delta\phi_{\rm tot} &=&\int \left[C_e \bar{Z}  n_{\rm tot} - \sum_{j,k} C_{j,k} n_{j,k}\right] dl \nonumber \\
&=& \int C_e \left[\bar{Z}  - \sum_{j,k} \frac{C_{j,k}}{C_e}\frac{n_{j,k}}{n_{\rm tot}}\right] n_{\rm tot} dl \nonumber \\
&\approx&  C_e [\bar{Z}  - Err] \int n_{\rm tot} dl,
\label{phase-err}
\end{eqnarray}
assuming a uniform $\bar{Z}$ along the path length through the plasma, and where 
\begin{equation}
Err = \sum_{j,k} \frac{C_{j,k}}{C_e}\frac{n_{j,k}}{n_{\rm tot}}.
\label{eq:err}
\end{equation}

If all the $C_{j,k}$ and $n_{j,k}$ in the plasma are known, then $Err$ can be 
calculated exactly. However, this is typically prohibitive due to a lack of complete information for both $C_{j,k}$ and
$n_{j,k}$. When $Err$ cannot
be calculated exactly, it is useful to determine bounds on $Err$
(and thus $n_{\rm tot}$). Using Eq.~(\ref{phase-err}) and $Err=0$ (i.e., only electrons present),
then the lower bound for $n_{\rm tot}$ is given by
\begin{eqnarray}
\left(\int n_{\rm tot} dl\right)_{\rm min} =  \frac{ \Delta\phi_{\rm tot}}{C_e \bar{Z} }.
\end{eqnarray}
Similarly, if we can determine the maximum $Err=Err_{\rm max}$, then an upper bound on $n_{\rm tot}$ is
given by
\begin{eqnarray}
\left(\int n_{\rm tot} dl\right)_{\rm max} = \frac{ \Delta\phi_{\rm tot}}{C_e [\bar{Z}  - Err_{\rm max}]}.
\end{eqnarray}
One method for determining $Err_{\rm max}$ is to determine $C_{\rm max}$ for all $j,k$ present
in the plasma, and then define $Err_{\rm max} \equiv C_{\rm max}/C_e$. 
Because $C_{\rm max} \ge C_{j,k}$ for all $j,k$ (by definition), then
\begin{eqnarray}
Err \le \frac{C_{\rm max}}{C_e} \sum_{j,k} \frac{n_{j,k}}{n_{\rm tot}} = \frac{C_{\rm max}}{C_e}=Err_{\rm max}
\end{eqnarray}
is always satisfied.
The problem then reduces to finding $C_{\rm max}$ for the given plasma. The $C_{j,k} = (2 \pi K_{j,k} m_k)/\lambda$, 
where $K_{j,k}$ is the Slater screening constant, $m_k$ is the mass,
and $\lambda$ is the interferometer laser wavelength. Since $K$ is 
proportional to the sum of mean square electron orbits for all bound electrons,\cite{alpher1} then for a given gas species 
$k$ the largest $K_{j,k}$ occurs for the neutral atom, i.e.,
$K_{\rm max} = K_{0,k}$. Thus, $C_{\rm max} = C_{0,k}$ 
for whichever gas species $k$ in the plasma has the largest neutral sensitivity constant. The maximum correction 
factor can be written as
\begin{eqnarray}
Err_{\rm max} &=& \frac{(C_{0,k})_{\rm max}}{C_e} = \frac{2 \pi }{C_e \lambda} (K_{0,k} m_k)_{\rm max}  \label{eq:err-max}
\end{eqnarray}
or, using $K_{j,k}m_k = (\delta N_n^{STP}/n_n^{STP})_k$,\cite{kumar,merritt-htpd12} 
\begin{eqnarray}
Err_{\rm max} &=& \frac{2 \pi }{C_e \lambda} \left( \frac{N_n^{STP} }{n_n^{STP}}\right)_{\rm k,max},
\end{eqnarray}
where $n_n^{STP}$ is the neutral density of the species at standard temperature and pressure, $N_n$ is the refractive index of the neutral species, $\delta N_n = N_n-1$, and $C_e = \lambda e^2/(4 \pi \epsilon_0 m_e c^2)$.

\section{Re-derivation of the Coulomb logarithm for counter-streaming ions in the presence of warm electrons}
\label{app:logL}

We point out an inconsistency in the 
Coulomb logarithm for counter-streaming ions with relative velocity $v_D=\beta_Dc$ in the presence of warm 
electrons ($kT_i/m_i,kT_{i^\prime}/m_{i^\prime} < v_D^2 < kT_e/m_e$), as given in the NRL Plasma
Formulary (2013 edition),\cite{nrl-formulary}
\begin{equation}
\lambda_{ii^\prime}=\lambda_{i^\prime i}= 35 - \ln\left[\frac{ZZ^\prime(\mu+\mu^\prime)}{\mu\mu^\prime
\beta_D^2}\left(\frac{n_e}{T_e}\right)^{1/2}\right],
\label{logL}
\end{equation}
where $T_e$ is in eV and units are cgs unless otherwise noted.
Unprimed and primed variables refer to test and field particles, respectively.
Equation~(\ref{logL}) affects ion collisionality estimates for counter-streaming
plasmas.\cite{drake12}

We re-derive the Coulomb logarithm using the definition employed in the NRL Plasma Formulary,\cite{nrl-formulary}
\begin{equation}
\lambda=\ln\Lambda=\ln\left(\frac{r_{\rm max}}{r_{\rm min}}\right),
\label{logL-definition}
\end{equation}
where in this case
\begin{equation}
r_{\rm max}=\lambda_{De}=\left(\frac{kT_e}{4\pi n_e e^2}\right)^{1/2}=
7.43\times10^2\left(\frac{T_e}{n_e}\right)^{1/2}
\label{lD}
\end{equation}
is the electron Debye length, and
$r_{\rm min}=ZZ^\prime e^2/(m_{ii^\prime} v_D^2)$
is the distance of closest approach between two counter-streaming ions
with reduced mass $m_{ii^\prime}=m_i m_{i^\prime}/(m_i+m_{i^\prime})$
and relative speed $v_D$.  We assume that $r_{\rm min}$
is greater than the de Broglie wavelength $\hbar/(2m_{ii^\prime} v_D)$.
We re-write $r_{\rm min}$ by pulling numerical constants to the front:
\begin{widetext}
\begin{equation}
r_{\rm min}=\frac{e^2}{m_p c^2} \frac{ZZ^\prime(\mu+\mu^\prime)}{\mu \mu^\prime (v_D/c)^2}=
\frac{(4.8032\times10^{-10})^2}{(1.6726\times 10^{-24})(2.9979\times10^{10})^2}\frac{ZZ^\prime
(\mu+\mu^\prime)}{\mu\mu^\prime \beta_D^2}=\\
1.5347\times 10^{-16}\frac{ZZ^\prime(\mu+\mu^\prime)}{\mu\mu^\prime\beta_D^2}.
\label{rmin}
\end{equation}
\end{widetext}
Substituting Eqs.~(\ref{lD}) and (\ref{rmin}) into Eq.~(\ref{logL-definition}), we obtain
\begin{widetext}
\begin{equation}
\lambda_{ii^\prime}=
\ln\Lambda=\ln\left\{\frac{743(T_e/n_e)^{1/2}}{1.5347\times10^{-16}[ZZ^\prime(\mu+\mu^\prime)]/(\mu\mu^\prime\beta_D^2)}\right\}
=43-\ln\left[\frac{ZZ^\prime (\mu+\mu^\prime)}{\mu\mu^\prime \beta_D^2}\left(\frac{n_e}{T_e}\right)^{1/2}\right]
\label{logL-correct}
\end{equation}
\end{widetext}
[same as Eq.~(\ref{logL-text})], which should supersede Eq.~(\ref{logL}).

The discrepancy between Eqs.~(\ref{logL-correct}) and (\ref{logL}) is exactly accounted for
if the constants $k=1.6022\times 10^{-12}$~erg/eV and $e^2=(4.8032\times10^{-10})^2$~statcoulomb$^2$ in 
Eq.~(\ref{lD}) are dropped.\cite{swadling-pc}
This seems like a plausible mistake to make in arriving at Eq.~(\ref{logL}).

For counter-streaming Al-Al collisions
with $\mu=\mu^\prime=27$, $Z=Z^\prime=2.0$, $v_D=20$~km/s, $T_e=2.2$~eV, and $n_e=6.5\times
10^{14}$~cm$^{-3}$ (corresponding to values in Table~II),
we calculate $\lambda_{ii} = 0.325$ and $\lambda_{ii}=8.3$ using Eqs.~(\ref{logL})
and (\ref{logL-correct}), respectively.  The latter is a more reasonable result for this weakly coupled example.

%

\end{document}